\documentclass[12pt]{article}
\usepackage{amsmath,amssymb}
\usepackage{amsthm}
\usepackage{bm}
\usepackage{color}
\usepackage{cite}
\usepackage{mathtools}
\usepackage{here}

\usepackage{hyperref}

\usepackage{multirow}
\usepackage {tabularx}
\newcolumntype{C}{>{\centering\arraybackslash}X}
\newcolumntype{L}{>{\raggedright\arraybackslash}X}
\newcolumntype{R}{>{\raggedleft\arraybackslash}X}

\numberwithin{equation}{section}

\begin{document}

\title{
\begin{flushright}
\ \\*[-80pt]
\begin{minipage}{0.25\linewidth}
\normalsize
EPHOU-24-012\\
KYUSHU-HET-293\\*[50pt]
\end{minipage}
\end{flushright}
{\Large \bf
Flavor symmetries from modular subgroups \\ in magnetized compactifications
\\*[20pt]}}

\author{
~Tatsuo Kobayashi$^1$, 
~Kaito Nasu$^1$, 
~Ryusei Nishida$^1$, \\
~Hajime Otsuka$^2$,   and
~Shohei Takada$^1$
\\*[20pt]
\centerline{
\begin{minipage}{\linewidth}
\begin{center}
$^{1}${\it \normalsize
Department of Physics, Hokkaido University, Sapporo 060-0810, Japan} \\*[5pt]
$^{2}${\it \normalsize
Department of Physics, Kyushu University, 744 Motooka, Nishi-ku, Fukuoka 819-0395, Japan} \\*[5pt]
\end{center}
\end{minipage}}
\\*[50pt]}

\date{
\centerline{\small \bf Abstract}
\begin{minipage}{0.9\linewidth}
\medskip
\medskip
\small
We study the flavor structures of zero-modes, which are originated from the modular symmetry on $T^2_1\times T^2_2$ and its orbifold with magnetic fluxes. We introduce the constraint on the moduli parameters by 
$\tau_2=N\tau_1$, where $\tau_i$ denotes the complex structure moduli on $T^2_i$. Such a constraint can be derived from the moduli stabilization. 
The modular symmetry of $T^2_1 \times T^2_2$ is $SL(2,\mathbb{Z})_{\tau_1} \times SL(2,\mathbb{Z})_{\tau_2} \subset Sp(4,\mathbb{Z})$ and it is broken to $\Gamma_0(N) \times \Gamma^0(N)$ by the moduli constraint. The wave functions represent their covering groups. We obtain various flavor groups in these models.
\end{minipage}
}

\begin{titlepage}
\maketitle
\thispagestyle{empty}
\end{titlepage}

\newpage


\section{Introduction}
\label{Intro}

It is one of the most important problems in particle physics to understand the origin of the flavor structure in three generations of quarks and leptons, 
i.e., hierarchical fermion masses and large and small mixing angles as well as CP phases.
Various approaches have been studied.
Among them, an introduction of flavor symmetries such as non-Abelian discrete symmetries as well as Abelian symmetries is 
one of the interesting approaches.
For example, $A_N$, $S_N$, $\Delta(3N^2)$, and $\Delta(6N^2)$ have been used as flavor symmetries in the bottom-up approach of flavor model building 
\cite{Altarelli:2010gt,Ishimori:2010au,Hernandez:2012ra,King:2013eh,Kobayashi:2022moq}.

Recently, modular flavor symmetries were studied intensively \cite{Feruglio:2017spp}.
The modular symmetry $SL(2,\mathbb{Z})$ is a geometrical symmetry of 
two-dimensional torus $T^2$ and its orbifold $T^2/\mathbb{Z}_2$.
Its finite modular groups $\Gamma_N$ include $S_3$, $A_4$, $S_4$, and $A_5$ \cite{deAdelhartToorop:2011re}.
Yukawa couplings and masses are written by modular forms in modular flavor symmetric models.
Modular forms of $S_3$\cite{Kobayashi:2018vbk}, $A_4$\cite{Feruglio:2017spp}, $S_4$ \cite{Penedo:2018nmg}, $A_5$\cite{Novichkov:2018nkm}, and their covering groups \cite{Liu:2019khw,Novichkov:2020eep,Liu:2020akv,Liu:2020msy} were studied.
One can construct realistic flavor models by use of these modular forms.
(See for reviews Refs.~\cite{Kobayashi:2023zzc,Ding:2023htn}.)

There appear flavor symmetries originated from the $SL(2,\mathbb{Z})$ modular symmetry in string compactifications such as heterotic orbifold models \cite{Ferrara:1989qb,Lerche:1989cs,Lauer:1989ax,Lauer:1990tm} and magnetized compactifications of type II string theory \cite{Kobayashi:2018rad,Kobayashi:2018bff,Ohki:2020bpo,Kikuchi:2020frp,Kikuchi:2020nxn,
Kikuchi:2021ogn,Almumin:2021fbk}. 
Moreover, generic string compactifications have more than one moduli field. 
They have larger symplectic modular symmetries, $Sp(2g,\mathbb{Z})$.
These modular flavor symmetries were studied in magnetized $T^{2g}$ 
and orbifold compactifications \cite{Kikuchi:2023awe,Kikuchi:2023dow}  and 
Calabi-Yau compactifications \cite{Strominger:1990pd,Candelas:1990pi,Ishiguro:2020nuf,Ishiguro:2021ccl} in the top-down approach.
They have a rich structure.
Also, these larger flavor symmetries were studied in model building of the bottom-up approach 
\cite{Ding:2020zxw,Ding:2021iqp,RickyDevi:2024ijc,Ding:2024xhz}.

On the other hand, subgroups of $SL(2,\mathbb{Z})$ and $Sp(2g,\mathbb{Z})$ also appear in string compactifications.
For example, the $\Gamma_0(N)$, $\Gamma^0(N)$, and their product groups appear in orbifold compactifications with certain lattices \cite{Mayr:1993mq,Ishiguro:2023wwf} 
and heterotic Calabi-Yau compactification 
in an asymptotic limit of moduli spaces \cite{Ishiguro:2024xph}.
They also appear in type IIB string theory, when the moduli space is constrained by 
the moduli stabilization due to three-form fluxes \cite{Kobayashi:2020hoc}.
Hence, it is interesting to study flavor symmetries originated from such subgroups.

Modular symmetries correspond to flavor symmetries of wave functions in magnetized compactifications 
\cite{Kobayashi:2018rad,Kobayashi:2018bff,Ohki:2020bpo,Kikuchi:2020frp,Kikuchi:2020nxn,
Kikuchi:2021ogn,Almumin:2021fbk}.
Thus, magnetized compactifications are very useful to understand flavor-symmetric aspects of the modular symmetries as a concrete setup. 
We consider $T^4$ and its orbifold compactifications with magnetic fluxes.
Our model may correspond to an effective field theory of magnetized $D7$-brane models. 
The geometrical modular symmetry of the complex structure moduli $\Omega$ on $T^4$ is $Sp(4,\mathbb{Z})$.
Wave functions represent their covering groups \cite{Kikuchi:2021ogn,Kikuchi:2023awe}.
For simplicity, we restrict ourselves to the factorizable $T_1^2\times T_2^2$, where 
only two complex structure moduli, $\tau_1$ and $\tau_2$, appear and 
the modular symmetry is described by $SL(2,\mathbb{Z})_{\tau_1} \times SL(2,\mathbb{Z})_{\tau_2} \subset Sp(4,\mathbb{Z})$.
Then, we introduce the constraint on the moduli as $\tau_2 = N\tau_1$.
Such a constraint can be realized by the superpotential induced by the three-form flux background \cite{Kobayashi:2020hoc}. 
That breaks the geometrical symmetry $SL(2,\mathbb{Z})_{\tau_1} \times SL(2,\mathbb{Z})_{\tau_2}$ 
to $\Gamma_0(N) \times \Gamma^0(N)$.
Under this constraint, we study the transformation behaviors of zero-mode wave functions, which correspond to flavor symmetries in these models.

The paper is organized as follows.
In section \ref{sec:modular}, we briefly review extra-dimensional models on $T^2$ and modular transformation behavior of their zero-modes.
In section \ref{sec:Flavor}, we study modular transformation behavior of zero-modes on $T^2_1\times T^2_2$ 
and its orbifold with magnetic fluxes under the condition $\tau_2 = N \tau_1$.
Section \ref{sec:conclusion} is devoted to the conclusion.
In Appendix \ref{app}, we give an explicit computation on the transformation behaviors of wave functions, which is used in section \ref{sec:Flavor}.


\section{Modular symmetry in magnetized compactifications}
\label{sec:modular}

Here, we give a brief review of zero-mode wave functions in torus and orbifold compactifications 
with magnetic fluxes and their transformation behaviors under the modular symmetry.

\subsection{Zero-mode wave functions}

First, we briefly review $T^2$ compactification with magnetic flux with an emphasis on zero-mode wave functions.
We use the complex coordinate $z=x+\tau y$, where $x$ and $y$ are real coordinates and $\tau$ is the 
complex structure modulus.
We identify $z \sim z + 1 \sim z+\tau$ on $T^2$.

We consider $U(1)$ gauge theory and introduce the magnetic flux background:
\begin{align}
F = \frac{\pi i M}{{\rm Im}\tau}dz \wedge d\bar z,
\end{align}
where $M$ must be an integer because of Dirac's quantization condition.
In what follows, we study the positive magnetic flux, i.e. $M>0$. 
Let us study zero-modes of the Dirac equation with $U(1)$ charge $q=1$ under the above magnetic flux 
background.
There appear $M$ zero-modes, and 
their wave functions $\psi^{j,M}(z,\tau)$ with $j=0,1,\cdots, M-1$ are written by \cite{Cremades:2004wa},
\begin{align}
\psi^{j,M}(z,\tau) = \frac{M^{1/4}}{{\cal A}^{1/2}}  e^{\pi iMz\frac{{\rm Im}z}{{\rm Im}\tau}}
  \vartheta \begin{bmatrix}
    \frac{j}{M} \\
    0 \\
  \end{bmatrix}
  (Mz,M\tau), \label{eq:psi} 
\end{align}
where $\cal A$ denotes the ares of $T^2$, and 
$\vartheta$ denotes the Jacobi theta function: 
\begin{align}
  \vartheta
  \begin{bmatrix}
    a \\ b \\
  \end{bmatrix}
  (\nu,\tau)
  = \sum_{\ell\in\mathbb{Z}} e^{\pi i(a+\ell)^2\tau} e^{2\pi i(a+\ell)(\nu+b)}.
\label{eq:theta-f}
\end{align}

The above zero-mode wave functions have the following property:
\begin{align}
\psi^{j,M}(z,\tau) =\psi^{M-j,M}(-z,\tau). 
\end{align}
Thus, one can construct zero-mode wave functions on $T^2/\mathbb{Z}_2$, which can be obtained by the identification $z\sim-z$ on $T^2$ \cite{Abe:2008fi}.
When $j \neq 0$ or $M/2$, 
even modes $\Psi^{j,M}_0(z,\tau)$ and odd modes $\Psi^{j,M}_1(z,\tau)$ on $T^2/\mathbb{Z}_2$ can be written by 
\begin{align}
\Psi^{j,M}_m(z,\tau) = \frac{1}{\sqrt 2}\left(\psi^{j,M}(z,\tau) +(-1)^m \psi^{M-j,M}(z,\tau) \right),
\end{align}
for $m=0,1$.
For $j=0$ and $M/2$, the zero-mode wave functions $\psi^{j,M}(z,\tau)$ also correspond to 
the even mode wave functions $\Psi^{j,M}_0(z,\tau)$ on $T^2/\mathbb{Z}_2$.

\subsection{Modular symmetry}

The modular group $\Gamma = SL(2,\mathbb{Z})$ is the group of $(2 \times 2)$ matrices:
\begin{align}
\gamma = 
\begin{pmatrix}
a & b\\ c & d\\
\end{pmatrix},
\end{align}
where $a,b,c,d$ are integers and satisfy $ad-bc=1$.
This group is generated by two generators, $S$ and $T$,
\begin{align}
S=
\begin{pmatrix}
0 & 1 \\ -1 & 0
\end{pmatrix}, 
\qquad 
T=
\begin{pmatrix}
1 & 1 \\ 0 & 1
\end{pmatrix}.
\end{align}
They satisfy 
\begin{align}
S^2=-1, \qquad (ST)^3=1.
\end{align}

The modulus $\tau$ transforms as 
\begin{align}
\tau \to \gamma \tau = \frac{a \tau +b}{c\tau +d},
\end{align}
while the coordinate $z$ also transforms as 
\begin{align}
z \to \gamma z=\frac{z}{c\tau +d}.
\end{align}

The zero-mode wave functions on $T^2$ transform under the double covering group of $\Gamma$, $\tilde \Gamma$ 
\cite{Kikuchi:2020frp,Kikuchi:2021ogn}.(See for $\tilde \Gamma$ Ref.~\cite{Liu:2020msy}.)
In order to represent $\tilde \Gamma$, we introduce a new element $\tilde Z$, which expands the center of $\Gamma$.
The generators $\tilde S$ and $\tilde T$ of $\tilde \Gamma$ satisfy the following algebraic relations:
\begin{align}
\tilde S^2 = \tilde Z, \quad (\tilde S \tilde T)=\tilde Z^2, \quad \tilde Z^4=1, \quad \tilde Z \tilde T = \tilde T \tilde Z.
\end{align}
Under the $\tilde S$ transformation, the zero-mode wave functions transform as 
\begin{align}
&\psi^{j,M}(\tilde S z, \tilde S\tau) =\tilde J(\tilde S,\tau)\sum_k \rho(\tilde S)_{jk}\psi^{k,M}(z,\tau), \notag \\
&\rho(\tilde S)_{jk}=e^{\pi i/4}\frac{1}{\sqrt M}e^{2\pi i \frac{jk}{M}}.
\end{align}
Here, $\tilde J(\tilde S,\tau)$ denotes the automorphy factor, and it is obtained as 
 $\tilde J(\tilde S,\tau)=(-\tau)^{1/2}$.\footnote{See also for discussions on modular weights Ref.~\cite{Kikuchi:2023clx}.}
That shows that the zero-mode wave functions have a modular weight $1/2$.
The $\tilde T$ transformation can be represented by the wave functions when $M=$ even \cite{Kikuchi:2021ogn}.\footnote{When we consider the wave functions with non-vanishing Sherk-Schwarz phases, the $\tilde{T}$ transformation can be represented even for $M=$ odd \cite{Kikuchi:2021ogn}.}
Thus, we focus on the case with $M=$ even.
Under the $\tilde T$ transformation, the zero-modes wave functions transform as 
\begin{align}
\label{eq:rho(T)}
&\psi^{j,M}(\tilde T z, \tilde T\tau) =\sum_k \rho(\tilde T)_{jk}\psi^{k,M}(z,\tau), \notag \\
&\rho(\tilde T)_{jk}=e^{\pi i\frac{j^2}{M}}\delta_{j,k}.
\end{align}
The elements $\rho(\tilde S)$ and $\rho(\tilde T)$ form a finite group \cite{Kikuchi:2021ogn}.
Note that $(\rho(\tilde{T}))^{2M}=1$, and it generates the $\mathbb{Z}_{2M}$ group.

Similarly, we can write the modular transformations in the $T^2/\mathbb{Z}_2$ basis.


\section{Flavor symmetries in magnetized compactifications}
\label{sec:Flavor}

In the previous section, we have reviewed the modular flavor symmetry of zero modes on $T^2$ and $T^2/\mathbb{Z}_2$.
One can extend it to $T^4$ and its orbifold, which may correspond to the effective field theory of magnetized $D7$-brane models.
In general, the modular symmetry of the complex structure moduli $\Omega$ of $T^4$ is 
$Sp(4,\mathbb{Z})$.
Here, we restrict ourselves to the factorizable 
$T_1^2\times T_2^2$ and its orbifold, where two complex structure moduli, $\tau_1$ and $\tau_2$, appear and their modular symmetry becomes $SL(2,\mathbb{Z})_{\tau_1} \times SL(2,\mathbb{Z})_{\tau_2} \subset Sp(4,\mathbb{Z})$.
We denote the coordinates on the first and second $T^2$ by $z_1$ and $z_2$, respectively.
Also, we introduce the magnetic fluxes $M_1$ and $M_2$ on $T^2_1$ and $T^2_2$, respectively, 
as in the previous section.
Then, zero-mode wave functions can be written by 
use of the wave functions $\psi^{j,M}(z,\tau)$  on $T^2$ in the previous section as $\psi^{j,M_1}(z_1,\tau_1)\psi^{k,M_2}(z_2,\tau_2)$.
Their flavor structure is a direct product of those on the first and second $T^2$'s.

The three-form flux background is one of definite moduli stabilization scenarios \cite{Gukov:1999ya}.
Certain three-form flux backgrounds lead to the relation between moduli as $\tau_2=N\tau_1\ (N\in\mathbb{N})$ \cite{Kobayashi:2020hoc} in the supersymmetric minimum. 
For example, when the three-form flux background induces the following superpotential:
\begin{align}
    W=(\tau_2-N\tau_1)g(\Phi),
\end{align}
where $\Phi$ denote moduli fields other than $\tau_1$ and $\tau_2$, 
the above constraint can be realized.
The modular subgroups such as $\Gamma_0(N)$, $\Gamma^0(N)$, and their products remain \cite{Kobayashi:2020hoc}.
In this section, we study the flavor symmetries of zero-modes when we constrain 
\begin{align}\label{eq:tau1-2}
\tau_2=N\tau_1.
\end{align}

\subsection{Subgroups}
\label{sec:subgroup}

First, we define congruence subgroups, $\Gamma_0(N)$ and $\Gamma^0(N)$ as follows,
\begin{align}
&\Gamma_0(N)=\left\{  
\begin{pmatrix}
a & b \\ c & d
\end{pmatrix} \in SL(2,\mathbb{Z}) ~\middle|~ c\equiv0~~~{\rm mod}~N
\right\}, \notag \\
&\Gamma^0(N)=\left\{  
\begin{pmatrix}
a & b \\ c & d
\end{pmatrix} \in SL(2,\mathbb{Z}) ~\middle|~ b\equiv0~~~{\rm mod}~N
\right\}.
\end{align}
Note that $\Gamma_0(1)=\Gamma^0(1)=\Gamma=SL(2,\mathbb{Z})$.

The modular transformation of $\tau_1$,
\begin{align}
\tau_1 \to \frac{a_1 \tau_1 +b_1}{c_1\tau_1 +d_1},
\end{align}
leads to the following transformation of $\tau_2$:
\begin{align}
\tau_2 \to \frac{a_1 \tau_2 +Nb_1}{\frac{c_1}{N}\tau_2 +d_1},
\end{align}
through the relation (\ref{eq:tau1-2}).
We may have a degree of freedom to multiply both the denominator and numerator by a common integer $m$, but 
$m$ must be $m= \pm 1$ by the following reason. 
We require that the following matrix:
\begin{align}
\label{eq:relation}
\begin{pmatrix}
a_2 & b_2 \\ c_2 & d_2
\end{pmatrix}=
\begin{pmatrix}
a_1 & Nb_1 \\ \frac{c_1}{N} & d_1
\end{pmatrix}
\end{align}
must belong to $SL(2,\mathbb{Z})$.
That is, we must have gcd$(a_2,b_2)={\rm gcd}(a_2,c_2)={\rm gcd}(b_2,d_2)={\rm gcd}(c_2,d_2)=1$, i.e., $m=\pm 1$.
Furthermore, the above requirement means that $c_1\equiv0$ mod $N$.
In addition, we find $b_2\equiv0$ mod $N$.
These results show that two $SL(2,\mathbb{Z})$ symmetries of $\tau_1$ and $\tau_2$ are broken down to
\begin{align}
\begin{pmatrix}
a_1 & b_1 \\ c_1 & d_1
\end{pmatrix} \in \Gamma_0(N),
\end{align}
and
\begin{align}
\begin{pmatrix}
a_2 & b_2 \\ c_2 & d_2
\end{pmatrix} \in \Gamma^0(N).
\end{align}
For example, the generators of $\Gamma_0(N)$ are written by Refs.~\cite{Rademacher,Chuman}:
\begin{align}
\Gamma_0(2)&\,:\,
    \begin{pmatrix}
      1 & 1\\
      0 & 1\\
    \end{pmatrix}=T,\quad 
    \begin{pmatrix}
      1 & 0\\
      2 & 1\\
    \end{pmatrix} = U^2,
\nonumber\\
\Gamma_0(3)&\,:\,
    \begin{pmatrix}
      1 & 1\\
      0 & 1\\
    \end{pmatrix}=T,\quad -
    \begin{pmatrix}
      1 & 0\\
      3 & 1\\
    \end{pmatrix}=S^2U^3,
\nonumber\\
\Gamma_0(4)&\,:\,
     \begin{pmatrix}
      -1 & 0\\
      0 & -1\\
    \end{pmatrix}=S^2, \quad   
  \begin{pmatrix}
      1 & 1\\
      0 & 1\\
    \end{pmatrix}=T,\quad 
    \begin{pmatrix}
      1 & 0\\
      4 & 1\\
    \end{pmatrix}=U^4,
\nonumber\\
\Gamma_0(5)&\,:\,
    \begin{pmatrix}
      1 & 1\\
      0 & 1\\
    \end{pmatrix}=T, \quad
    \begin{pmatrix}
      1 & 0\\
      5 & 1\\
    \end{pmatrix}=U^5, \quad 
    \begin{pmatrix}
      3 & 1\\
      5 & 2\\
    \end{pmatrix}=U^2SU^3,
\end{align}
where $U=S^2TST$. 
Similarly, the generators of $\Gamma^0(N)$ are written by
\begin{align}
\Gamma^0(2)&\,:\,    
    \begin{pmatrix}
      1 & 0\\
      1 & 1\\
    \end{pmatrix}=U,\quad 
    \begin{pmatrix}
      1 & 2\\
      0 & 1\\
    \end{pmatrix} = T^2,
\nonumber\\
\Gamma^0(3)&\,:\,    
    \begin{pmatrix}
      1 & 0\\
      1 & 1\\
    \end{pmatrix}=U, \quad-
    \begin{pmatrix}
      1 & 3\\
      0 & 1\\
    \end{pmatrix}=S^2T^3,
\nonumber\\
\Gamma^0(4)&\,:\,    
     \begin{pmatrix}
      -1 & 0\\
      0 & -1\\
    \end{pmatrix}=S^2, \quad   
  \begin{pmatrix}
      1 & 0\\
      1& 1\\
    \end{pmatrix}=U,\quad 
    \begin{pmatrix}
      1 & 4\\
      0 & 1\\
    \end{pmatrix}=T^4,
\nonumber\\
\Gamma^0(5)&\,:\,    
    \begin{pmatrix}
      1 & 0\\
      1 & 1\\
    \end{pmatrix}=U, \quad
    \begin{pmatrix}
      1 & 5\\
      0 & 1\\
    \end{pmatrix}=T^5, \quad 
    \begin{pmatrix}
      3 & 5\\
      1 & 2\\
    \end{pmatrix}=T^3S^3T^2.
\end{align}
Note the correspondence of generators between 
$\Gamma_0(N)$ and $\Gamma^0(N)$ through Eq.~(\ref{eq:relation}). 
Specifically, $\Gamma^0(N)$ and $\Gamma_0(N)$ are  isomorphic to each other, and the generators of $\Gamma^0(N)$ are given by the transposed generators of $\Gamma_0(N)$.

\subsection{Symmetries of zero-modes}

Now, let us study the transformation behaviors of zero-mode wave functions 
under the constraint $\tau_2=N\tau_1$.
At the moment, we study transformation behaviors of wave functions only on the first 
$T_1^2$. 
Since we have already reviewed their transformation behaviors under $\tilde S$ and 
$\tilde T$, we focus on transformations by $\tilde{U}=\tilde{S}^2\tilde{T}\tilde{S}\tilde{T}$.
In general, the zero-mode wave functions on $T_1^2$ transform under $\tilde U^N$ as
\begin{align}
&\psi^{j,M_1}(\tilde{U}^Nz_1, \tilde{U}^N\tau_1) = \tilde{J}(\tilde{U}^N, \tau_1)\sum_{k}\rho(\tilde{U}^N)_{jk}\psi^{k,M_1}(z_1, \tau_1),
\end{align}
where
\begin{align}
& \tilde{J}(\tilde{U}^N, \tau_1)=(N\tau_1+1)^{1/2},  \\
&\rho(\tilde{U}^N)_{jk} = \frac{(-1)^{N-1}i}{\sqrt{NM_1}}e^{\pi i/4}e^{\frac{\pi i}{NM_1}(j-k)^2}\sum^{N-1}_{l = 0}e^{\frac{\pi i M_1}{N}l^2}e^{-\frac{2\pi i}{N}(j-k)l}. \notag 
\end{align}
The explicit computation of $\rho(\tilde{U}^N)$ is given in Appendix A.
Then, the zero-mode wave functions $\Psi^{j,M}_{m}$ on $T^2/\mathbb{Z}_2$ transform as 
\begin{align}
    &\Psi^{j,M}_{m}(\tilde{\gamma}z, \tilde{\gamma}\tau) = \tilde{J}(\tilde{\gamma}, \tau)\sum^{M/2}_{k=0}\rho_{m}(\tilde{\gamma})_{jk}\Psi^{k, M}_{m}(z, \tau), \notag \\
    &\rho_{0}(\tilde{S})_{jk} = \mathcal{N}^{j}_{t,2}\mathcal{N}^{k}_{t,2}\frac{4e^{\pi i/4}}{\sqrt{M}}\cos{\Big(\frac{2\pi jk}{M}\Big)}, \quad
    \rho_{0}(\tilde{T})_{jk} = e^{\pi i\frac{j^2}{M}}\delta_{j,k} ~~{\rm for }~~{\rm even \ modes}, \notag\\
    &\rho_{1}(\tilde{S})_{jk} = \mathcal{N}^{j}_{t,2}\mathcal{N}^{k}_{t,2}\frac{4ie^{\pi i/4}}{\sqrt{M}}\sin{\Big(\frac{2\pi jk}{M}\Big)}, \quad
    \rho_{1}(\tilde{T})_{jk} = e^{\pi i\frac{j^2}{M}}\delta_{j,k}~~{\rm for }~~{\rm odd \ modes},
\end{align}
where
\begin{align}
      \mathcal{N}^{j}_{t,2} = 
  \left\{ \,
    \begin{aligned}
    & 1/2 \quad (j = 0,M/2)\\
    & 1/\sqrt{2}\quad (\rm otherwise) \\
    \end{aligned}
\right. 
.
\end{align}

\subsubsection{Models with $M_1 \neq 0$ and $M_2=0$}

First, let us consider the case of $M_1 = 2$, where  two zero-modes 
$\psi^{j,2}(z_1,\tau_1)$ $(j=0,1)$ appear on $T^2_1$.
Note that both of them are also even modes on $T^2_1/\mathbb{Z}_2$.
For the second $T^2_2$, we set $M_2=0$.
On the orbifold $T^2_2/\mathbb{Z}_2$, there is a single even zero-mode, while 
odd-modes have no zero-mode.
Note that the wave function of the even zero-mode is constant, and its modular transformation is trivial.
On this orbifold $T^2_1/\mathbb{Z}_2 \times T^2_2/\mathbb{Z}_2$, the flavor symmetry 
is determined only by $\psi^{j,2}(z_1,\tau_1)$ $(j=0,1)$.

In this case, $\rho(\tilde{U}), \rho(\tilde{U}^2), \rho(\tilde{U}^3)$ and $\rho(\tilde{U}^4)$ are represented as 
\begin{align}
    &\rho(\tilde{U}) = \frac{e^{\pi i/4}}{\sqrt{2}}
    \begin{pmatrix}
        i & -1\\
        -1 & i\\
    \end{pmatrix}, \quad
    \rho(\tilde{U}^2) =
    \begin{pmatrix}
        0 & 1\\
        1 & 0\\
    \end{pmatrix}, \notag\\
    &\rho(\tilde{U}^3) = 
    \frac{e^{\pi i/4}}{\sqrt{2}}
    \begin{pmatrix}
        -1 & i\\
        i & -1\\
    \end{pmatrix}, \quad
    \rho(\tilde{U}^4) = 
    \begin{pmatrix}
        1 & 0\\
        0 & 1\\
    \end{pmatrix},
\end{align}
by the two zero-modes on the first $T^2_1$.
Thus, we find  $ \rho(\tilde{U})$ as well as $ \rho(\tilde{U}^{2n+1})$ generate
$\mathbb{Z}_4$, while $\rho(\tilde{U}^2)$ as well as $\rho(\tilde{U}^{4n+2})$
generate $\mathbb{Z}_2$. 
That is reasonable as follows.
The generator $\rho(\tilde{T})$ satisfies $(\rho(\tilde{T}))^{2M}=1$ by Eq.~(\ref{eq:rho(T)}), i.e., $\mathbb{Z}_{2M}$.
When $M=2$, the generator $\rho(\tilde{T})$ generates the $\mathbb{Z}_4$ group.
The generator $U$ of $\Gamma^0(N)$ is a kind of dual of $T$, which generates $\Gamma_0(N)$.
They satisfy the same relation, $\rho(\tilde{T}^4)=\rho(\tilde{U}^4)=1$ on the wave functions.

Also, $\rho(\tilde{S}^2\tilde{U}^3)$ generates $\mathbb{Z}_4$ because of

\begin{align}
    &\rho(\tilde{S}^2\tilde{U}^3) = \frac{e^{\pi i/4}}{\sqrt{2}}
    \begin{pmatrix}
        -i & -1\\
        -1 & -i\\ 
    \end{pmatrix}, \quad
    \rho(\tilde{S}^2\tilde{U}^3)^2 =
    \begin{pmatrix}
        0 & -1\\
        -1 & 0\\
    \end{pmatrix}, \notag\\
    &\rho(\tilde{S}^2\tilde{U}^3)^3 = \frac{e^{\pi i/4}}{\sqrt{2}}
    \begin{pmatrix}
        1 & i\\
        i & 1\\
    \end{pmatrix}, \quad
    \rho(\tilde{S}^2\tilde{U}^3)^4 =
    \begin{pmatrix}
        1 & 0\\
        0 & 1\\
    \end{pmatrix}.
\end{align}

The total flavor group includes the $\mathbb{Z}_4$ of $\rho(\tilde{T})$ and 
$\mathbb{Z}_4$ or $\mathbb{Z}_2$ of $\rho(\tilde{U})$.
By use of the above results, we can identify the full symmetries of 
zero-modes on the first $T_1^2$ with $\tau_2=N \tau_1$.
For $N=2$, the generators,
\begin{align}
    &\rho(\tilde{T}) = 
    \begin{pmatrix}
        1 & 0\\
        0 & i\\
    \end{pmatrix},\quad
    \rho(\tilde{U}^2)= 
    \begin{pmatrix}
        0 & 1\\
        1 & 0\\
    \end{pmatrix},
\end{align}
generate $\Sigma(32)$, which is a non-Abelian group 
because $\rho(\tilde{T})$ and $\rho(\tilde{U}^2)$ are not commutable.

For $N=3$, the generators,
\begin{align}
   \rho(\tilde{T}) =  \begin{pmatrix}
        1 & 0\\
        0 & i\\
    \end{pmatrix},\quad
    \rho(\tilde{S}^2\tilde{U}^3)= \frac{e^{\pi i/4}}{\sqrt{2}}
    \begin{pmatrix}
        -i & -1\\
        -1 & -i\\
    \end{pmatrix},
\end{align}
generate $T'\rtimes \mathbb{Z}_4$, which is a non-Abelian group similar to the above.

For $N=4$, the generators,
\begin{align}
    \rho(\tilde{S}^2) = i
    \begin{pmatrix}
        1 & 0\\
        0 & 1\\
    \end{pmatrix},\quad
    \rho(\tilde{T}) = 
    \begin{pmatrix}
        1 & 0\\
        0 & i\\
    \end{pmatrix},\quad
    \rho(\tilde{U}^4) = 
    \begin{pmatrix}
        1 & 0\\
        0 & 1\\
    \end{pmatrix},
\end{align}
generate $\mathbb{Z}_4 \times \mathbb{Z}_4$, which is an Abelian group because all of three generators, $\rho(\tilde{S}^2)$, $\rho(\tilde{T})$, and $\rho(\tilde{U}^4)$, are represented by diagonal matrices.
In particular,  $\rho(\tilde{U}^4)$ is the identity as studied above.

For $N=5$, the generators,
\begin{align}
    \rho(\tilde{T}) = 
    \begin{pmatrix}
        1 & 0\\
        0 & i\\
    \end{pmatrix},\quad
    \rho(\tilde{U}^5) = \frac{e^{\pi i/4}}{\sqrt{2}}
    \begin{pmatrix}
        i & -1\\
        -1 & i\\
    \end{pmatrix},\quad
    \rho(\tilde{U}^2\tilde{S}\tilde{U}^3)= \frac{e^{\pi i/4}}{\sqrt{2}}
    \begin{pmatrix}
        -i & i\\
        -1 & -1\\
    \end{pmatrix},
\end{align}
generate $T'\rtimes \mathbb{Z}_4$.
Results are summarized in Table \ref{M=2,Gamma_0(N)}.
The zero-modes $\psi^{j,2}$ ($j=0,1$) are doublets of these flavor groups, although the doublet is decomposed to two singlets under the Abelian $\mathbb{Z}_4 \times \mathbb{Z}_4$ for $N=4$.

Note that the same group $T'\rtimes \mathbb{Z}_4$ appears 
when $N=3$ and $N=5$. 
We have shown that $\rho(\tilde U^4)$ is identity.
That leads to $\rho(\tilde U^5)=\rho(\tilde{U})$ on the wave functions.
In addition, we find
\begin{align}    \rho(\tilde{S})=\left(\rho(\tilde{T}^{-1})\rho(\tilde{U})
    \rho(\tilde{T}^{-1})\right)^3.
\end{align}
That implies that the group generated by 
$\rho(\tilde{T})$ and $\rho(\tilde{S})$ is the same as one generated by $\rho(\tilde{T})$ and $\rho(\tilde{U})$.
Indeed when $N=1$, we obtain the same group $T'\rtimes \mathbb{Z}_4$ \cite{Kikuchi:2020frp}.
Similarly, we can discuss the cases with $N=$ odd.
On the other hand, when $N=2$, the flavor group becomes smaller, but non-Abelian.
Moreover, when $N=4$, the flavor group becomes much smaller, i.e., 
 Abelian.

For $N=2$, we examine the transformation behavior of 
$(\psi^{j,2})^4$ $(j=0,1)$.
They have the modular weight 2 because $\psi^{j,2}$ themselves have the modular weight 1/2.
Also $(\psi^{j,2})^4$ are invariant under $\Sigma(32)$.
That is consistent with the fact that 
$\Gamma_0(N)$ has two singlet modular forms of the weight 2.
(See for review on modular forms of $\Gamma_0(N)$ Ref.~\cite{DHoker:2022dxx}.)
Similarly, for other $N$, $(\psi^{j,2})^4$ include 
the modular forms of the weight 2, which are invariant under 
the flavor groups.

\begin{table}[H]
    \centering
    \begin{tabular}{|c|c|c|}
  \hline
  $N$ & order & flavor group\\
  \hline
    \hline
    $2$& 32& $\Sigma(32)$\\
    \hline
    $3$& 96& $T' \rtimes\mathbb{Z}_4$\\
    \hline
    $4$& 16& $\mathbb{Z}_4 \times \mathbb{Z}_4$\\
    \hline
    $5$& 96& $T' \rtimes \mathbb{Z}_4$\\
    \hline
    \end{tabular}
    \caption{The flavor groups for $M_1=2$.}
    \label{M=2,Gamma_0(N)}
\end{table}

Similarly, we can discuss the case with $M_1=4$ and $M_2=0$ under the condition $\tau_2=N\tau_1$.
The flavor structure on 
$T^2_1/\mathbb{Z}_2 \times T^2_2/\mathbb{Z}_2$ is determined by the first $T^2_1/\mathbb{Z}_2$.
The number of even zero-modes is three, while the number of odd zero-modes is one.
Thus, the even modes can correspond to three generations.
The $\rho(\tilde{T})$ generator satisfies $(\rho(\tilde{T}))^{2M}=(\rho(\tilde{T}))^{8}=1$, i.e., $\mathbb{Z}_8$.
On the other hand, the generator $\rho(\tilde{U})$ is 
represented by
\begin{align}
    \rho(\tilde{U})=  \frac{ie^{\pi i/4}}{2}
    \begin{pmatrix}
        1 & e^{\pi i/4} & -1 & e^{\pi i/4}\\
        e^{\pi i/4} & 1 & e^{\pi i/4} & -1\\
        -1 & e^{\pi i/4} & 1 & e^{\pi i/4}\\
        e^{\pi i/4} & -1 & e^{\pi i/4} & 1\\
    \end{pmatrix}.
\end{align}
It satisfies $(\rho(\tilde{U}))^{8}=1$, i.e., $\mathbb{Z}_8$.
They satisfy the same relation, $(\rho(\tilde{T}))^8=(\rho(\tilde{U}))^8=1$.
That is reasonable as mentioned above.
Thus, the total flavor group includes the 
$\mathbb{Z}_8$ of $\rho(\tilde{T})$ and 
$\mathbb{Z}_8$ or its subgroups of $\rho(\tilde{U})$.
The flavor symmetries of these three generations are 
obtained as follows.

For $N=2$, the generators are represented by
\begin{align}
    &\rho_{0}(\tilde{T}) = 
    \begin{pmatrix}
        1 & 0 & 0\\
         0 & e^{\pi i/4} & 0\\
         0 & 0 & -1\\
    \end{pmatrix},\quad
    \rho_{0}(\tilde{U}^2) = \frac{-ie^{\pi i/4}}{\sqrt{2}}
    \begin{pmatrix}
    1 & 0 & i\\
    0 & \sqrt{2}e^{\pi i/4} & 0\\
    i & 0 & 1\\
    \end{pmatrix}.
\end{align}
They are reducible and can be decomposed into 
\begin{align}
    \rho_{0}(\tilde{T}) = 
    \begin{pmatrix}
        1 & 0\\
        0 & -1\\
    \end{pmatrix},\quad
    \rho_{0}(\tilde{U}^2) = \frac{-ie^{\pi i/4}}{\sqrt{2}}
    \begin{pmatrix}
        1 & i\\
        i & 1\\
    \end{pmatrix},
\end{align}
and
\begin{align}
     \rho_{0}(\tilde{T})= e^{\pi i/4},\quad \rho_{0}(\tilde{U}^2) = 1.
\end{align}
The former is the doublet of $\Sigma(32)$, while the latter is the $\mathbb{Z}_8$ group. The former group is non-Abelian because the generators $\rho_0(\tilde{T})$ and $\rho_0(\tilde{U}^2)$ are not commutable.

For $N=3$, the generators
\begin{align}
    \rho_{0}(\tilde{T}) = 
    \begin{pmatrix}
        1 & 0 & 0\\
         0 & e^{\pi i/4} & 0\\
         0 & 0 & -1\\
    \end{pmatrix},\quad 
    \rho_{0}(\tilde{S}^2\tilde{U}^3) = \frac{ie^{\pi i/4}}{2}
    \begin{pmatrix}
        1 & \sqrt{2}ie^{\pi i/4} & -1\\
        \sqrt{2}ie^{\pi i/4} & 0 & \sqrt{2}ie^{\pi i/4}\\
        -1 & \sqrt{2}ie^{\pi i/4} & 1\\
    \end{pmatrix},
\end{align}
consist the group $\Delta(48)\rtimes \mathbb{Z}_8$.

For $N=4$, the generators are represented by 
\begin{align}
    \rho_{0}(\tilde{S}^2) = i
    \begin{pmatrix}
        1 & 0 & 0\\
         0 & 1 & 0\\
         0 & 0 & 1\\
    \end{pmatrix},\quad
    \rho_{0}(\tilde{T}) = 
    \begin{pmatrix}
        1 & 0 & 0\\
         0 & e^{\pi i/4} & 0\\
         0 & 0 & -1\\
    \end{pmatrix},\quad 
    \rho_{0}(\tilde{U}^4) = 
    \begin{pmatrix}
        0 & 0 & 1\\
        0 & 1 & 0\\
        1 & 0 & 0\\
    \end{pmatrix}.
\end{align}
They are reducible and can be decomposed to 
\begin{align}
    \rho_{0}(\tilde{S}^2) = i
    \begin{pmatrix}
        1 & 0 \\
          0 & 1\\
    \end{pmatrix},\quad
    \rho_{0}(\tilde{T}) = 
    \begin{pmatrix}
        1 & 0 \\
         0 & -1\\
    \end{pmatrix},\quad 
    \rho_{0}(\tilde{U}^4) = 
    \begin{pmatrix}
        0 & 1\\
        1 & 0 \\
    \end{pmatrix},
\end{align}
and 
\begin{align}
    \rho_{0}(\tilde{S}^2) = i,\quad
    \rho_{0}(\tilde{T}) = e^{\pi i /4},\quad 
    \rho_{0}(\tilde{U}^4) = 1.
\end{align}
The former is the doublet of $D_4 \rtimes \mathbb{Z}_2$, 
while the latter is the $\mathbb{Z}_8$ group.
The former flavor group is non-Abelian because the generators $\rho(\tilde{T})$ and $\rho(\tilde{U}^4)$ are not commutable.
That is different from the case with $M_1=2$.

For $N=5$, the generators
\begin{align}
    &\rho_{0}(\tilde{T}) = 
    \begin{pmatrix}
        1 & 0 & 0\\
         0 & e^{\pi i/4} & 0\\
         0 & 0 & -1\\
    \end{pmatrix},\quad 
    \rho_{0}(\tilde{U}^5) = \frac{ie^{\pi i/4}}{2}
    \begin{pmatrix}
        -1 & \sqrt{2}e^{\pi i/4} & 1\\
        \sqrt{2}e^{\pi i/4} & 0 & \sqrt{2}e^{\pi i/4}\\
        1 & \sqrt{2}e^{\pi i/4} & -1\
    \end{pmatrix}
    ,\quad\notag\\
    &\rho_{0}(\tilde{U}^2\tilde{S}\tilde{U}^3) = \frac{e^{\pi i/4}}{2}
    \begin{pmatrix}
        i & -\sqrt{2} & i\\
        \sqrt{2}e^{\pi i/4} & 0 & -\sqrt{2}e^{\pi i/4}\\
        -i & -\sqrt{2} & -i\\
    \end{pmatrix},
\end{align}
generate the group $\Delta(48)\rtimes \mathbb{Z}_8$.
Results are summarized in Table \ref{M=4,Gamma_0(N)}.
The even zero-modes $\Psi^{j,4}_0$ are triplets under these flavor groups. 
For $N=2$ and 4, the triplet is reducible and decomposed into the doublet and singlet.

Similar to $M_1=2$, both the cases with $N=3,5$ lead to 
the same group $\Delta(48)\rtimes \mathbb{Z}_8$.
The same group is also obtained when $N=1$ \cite{Kikuchi:2020frp}.
For $N=2$, the flavor group is smaller, but non-Abelian.
For $N=4$, the flavor group is much smaller but non-Abelian.

For $N=2$, we examine the transformation behavior of 
$(\Psi^{j,4}_0)^4$ corresponding to the doublet.
They have the modular weight 2 and are invariant under $\Sigma(32)$.
That is consistent with the modular forms of $\Gamma_0(N)$, 
similar to the case with $M_1=2$.
For other $N$, we find the same aspect.

\begin{table}[H]
    \centering
    \begin{tabular}{|c|c|c|}
    \hline
         $N$ & order& flavor structure  \\
         \hline
         \hline
         $2$& 32,8& doublet of $\Sigma(32)$  and singlet of $\mathbb{Z}_8$\\
         \hline
         $3$& 384&  triplet of $\Delta(48)\rtimes \mathbb{Z}_8$\\
         \hline
         $4$&16,8 & doublet of $D_4 \rtimes \mathbb{Z}_2$ and singlet of $\mathbb{Z}_8$ \\
         \hline
         $5$& 384& triplet of $\Delta(48)\rtimes \mathbb{Z}_8$\\
         \hline

    \end{tabular}
    \caption{The flavor groups of three $\mathbb{Z}_2$ even zero-modes for $M_1=4$. For $N=2$ and 4, the triplet is reducible and decomposed into the doublet and singlet. The flavor groups of the doublet and the singlet are shown.}
    \label{M=4,Gamma_0(N)}
\end{table}

We also study the flavor groups for the odd zero-mode on $T^2_1/\mathbb{Z}_2$ with $M_1=4$, whose number is just one.
The generators are written by 
\begin{align}
    &\rho_1(\tilde{T}) = e^{\pi i/4} ,\quad
    \rho_{1}(\tilde{U}^2) = e^{\pi i/2}
    ~~{\rm for }~~N=2, \notag\\
    &\rho_1(\tilde{T}) = e^{\pi i/4} ,\quad
    \rho_{1}(\tilde{S}^2\tilde{U}^3) = -e^{\pi i/4} ~~{\rm for }~~N=3, \notag \\
    &\rho_1(\tilde{S}^2) = i ,\quad
    \rho_1(\tilde{T}) = e^{\pi i/4},\quad
    \rho_{1}(\tilde{U}^4) = -1~~{\rm for }~~N=4,\\
    &\rho_1(\tilde{T}) = e^{\pi i/4},\quad
    \rho_{1}(\tilde{U}^5) = -e^{\pi i/4},\quad
    \rho_{1}(\tilde{U}^2\tilde{S}\tilde{U}^3) = -i~~{\rm for }~~N=5. \notag
\end{align}
All of them generate the $\mathbb{Z}_8$ flavor group.

\subsubsection{Models with $M_1,M_2 \neq 0$}

We move to the case that the wave functions have non-trivial profiles on both $T^2_1$ and $T^2_2$ with $M_1 \neq 0$ and 
$M_2 \neq 0$.
In general, the zero-mode wave functions on $T^2_1\times T^2_2$ behave under the modular transformation as
\begin{align}
    &\psi^{j,M_1}_{T^2_1}(\tilde{\gamma}_1z_1, \tilde{\gamma}_1\tau_1 )\psi^{k,M_2}_{T^2_2}(\tilde{\gamma}_2z_2, \tilde{\gamma}_2\tau_2 ) = J_1(\gamma_1,\tau_1)\sum_{m,n}\rho(\tilde{\gamma}_1)_{jm}\rho(\tilde{\gamma}_2)_{kn}\psi^{m,M_1}_{T^2_1}(z_1,\tau_1)\psi^{n,M_2}_{T^2_{2}}(z_2,\tau_2), \notag \\
    &\rho_{T^2_1\times T^2_2}(\tilde{T}, \tilde{T}^N)_{(jk)(mn)} = \rho_{T^2_1}(\tilde{T})_{jm}\rho_{T^2_2}(\tilde{T}^N)_{kn} = 
    e^{\pi i(\frac{j^2}{M_1}+\frac{Nk^2}{M_2})}\delta_{j,m}\delta_{k,n}, \notag \\
    &\rho_{T^2_1\times T^2_2}(\tilde{U}^N, \tilde{U})_{(jk)(mn)} = \rho_{T^2_1}(\tilde{U}^N)_{jm}\rho_{T^2_2}(\tilde{U})_{kn} = 
    \frac{i(-1)^Ne^{\pi i(\frac{(j-m)^2}{NM_1} + \frac{(k-n)^2}{M_2})}}{\sqrt{NM_1M_2}}\sum^{N-1}_{l=0}e^{\frac{\pi iM_1}{N}l^2}e^{-\frac{2\pi i}{N}(j-m)l}.
\end{align}
Note that $J_1(\gamma_1,\tau_1)$ is an automorphy factor of the modular forms with weight $1$, and that $\tilde \gamma_1$ and $\tilde \gamma_2$ must satisfy 
the relation (\ref{eq:relation}).

First, we discuss the three-generation model on $T^2_1/\mathbb{Z}_2 \times T^2_2/\mathbb{Z}_2$.
The three zero-modes can be realized for a combination of 
even modes on $T^2_1/\mathbb{Z}_2$ with $M_1=4$ 
and odd mode on $T^2_2/\mathbb{Z}_2$ with $M_2=4$.
In this case, the zero-mode wave functions are obtained by 
$\Psi^{j,4}_{T^2_1/\mathbb{Z}_20}\Psi^{k,4}_{T^2_2/\mathbb{Z}_21} $.
Their transformation behaviors are given by\ 
$\Psi^{j,4}_{T^2_1/\mathbb{Z}_20}\Psi^{k,4}_{T^2_2/\mathbb{Z}_21} $  as 
\begin{align}
    &\Psi^{j,4}_{T^2_1/\mathbb{Z}_20}(\tilde{\gamma}_1z_1,\tilde{\gamma}_1\tau_1)\Psi^{k,4}_{T^2_2/\mathbb{Z}_21}(\tilde{\gamma}_2z_2,\tilde{\gamma}_2\tau_2) \notag \\ &= J_1(\gamma_1,\tau_1)\sum_{m,n}\rho_0(\tilde{\gamma}_1)_{jm}\rho_1(\tilde{\gamma}_2)_{kn}\Psi^{m,4}_{T^2_1/\mathbb{Z}_20}(z_1,\tau_1)\Psi^{n,4}_{T^2_2/\mathbb{Z}_21}(z_2,\tau_2)\notag\\
    &= J_1(\gamma_1,\tau_1)\sum_{m,n}\rho_{0,1}(\tilde{\gamma}_1,\tilde{\gamma}_2)_{(jk)(mn)}\Psi^{m,4}_{T^2_1/\mathbb{Z}_20}(z_1,\tau_1)\Psi^{n,4}_{T^2_2/\mathbb{Z}_21}(z_2,\tau_2).
\end{align}
Note that we have the correspondence between 
$\tilde \gamma_1$ and $\tilde \gamma_2$ through 
Eq.~(\ref{eq:relation}).

For $N=2$, the generators are represented by 
\begin{align}
    & \rho_{0,1}(\tilde{T},\tilde{T}^2) = i
    \begin{pmatrix}
        1 & 0 & 0\\
        0 & e^{\pi i/4} & 0\\
        0 & 0 & -1\\
    \end{pmatrix},\quad 
    \rho_{0,1}(\tilde{U}^2, \tilde{U}) = \frac{-1}{\sqrt{2}}
    \begin{pmatrix}
    1 & 0 & i\\
    0 & \sqrt{2}e^{\pi i/4} & 0\\
    i & 0 & 1\\
    \end{pmatrix}.
    \end{align}
    They are reducible and can be decomposed into 
    \begin{align}
        \rho_{0,1}(\tilde{T},\tilde{T}^2) = 
        \begin{pmatrix}
            i & 0\\
            0  & -i\\
        \end{pmatrix},\quad
        \rho_{0,1}(\tilde{U}^2,\tilde{U}) = \frac{-1}{\sqrt{2}}
        \begin{pmatrix}
            1 & i\\
            i &1\\
        \end{pmatrix},
    \end{align}
    and
    \begin{align}
        \rho_{0,1}(\tilde{T},\tilde{T}^2) = ie^{\pi i/4},\quad 
        \rho_{0,1}(\tilde{U}^2,\tilde{U}) = -e^{\pi i/4}.
    \end{align}
    The former is the doublet of $Q_{8}$, while the latter is the $\mathbb{Z}_8$ group.

For $N=3$, the generators 
    \begin{align}
    & \rho_{0,1}(\tilde{T},\tilde{S}^2\tilde{T}^3) = ie^{3\pi i/4}
    \begin{pmatrix}
        1 & 0 & 0\\
        0 & e^{\pi i/4} & 0\\
        0 & 0 & -1\\
    \end{pmatrix},\notag\\
    &\rho_{0,1}(\tilde{S}^2\tilde{U}^3,\tilde{U}) =\frac{1}{2}
    \begin{pmatrix}
        1 & \sqrt{2}ie^{\pi i/4} & -1\\
        \sqrt{2}ie^{\pi i/4} & 0 & \sqrt{2}ie^{\pi i/4}\\
        -1 & \sqrt{2}ie^{\pi i/4} & 1\\
    \end{pmatrix},
\end{align}
generate the flavor group $\Delta(48)\rtimes (\mathbb{Z}_2\times \mathbb{Z}_2)$.

For $N=4$, the generators are represented by 
\begin{align}
    & \rho_{0,1}(\tilde{S}^2,\tilde{S}^2) = -
    \begin{pmatrix}
        1 & 0 & 0\\
        0 & 1 & 0\\
        0 & 0 & 1\\
    \end{pmatrix},\quad
    \rho_{0,1}(\tilde{T},\tilde{T}^4) = -
    \begin{pmatrix}
        1 & 0 & 0\\
         0 & e^{\pi i/4} & 0\\
         0 & 0 & -1\\
    \end{pmatrix},\notag\\
    &\rho_{0,1}(\tilde{U}^4,\tilde{U}) = ie^{3\pi i/4}
    \begin{pmatrix}
        0 & 0 & 1\\
        0 & 1 & 0\\
        1 & 0 & 0\\
    \end{pmatrix}.
    \end{align}
    They are reducible and can be decomposed into 
    \begin{align}
        \rho_{0,1}(\tilde{S}^2,\tilde{S}^2) = - 
        \begin{pmatrix}
            1 & 0\\
            0 & 1\\
        \end{pmatrix},\quad
        \rho_{0,1}(\tilde{T},\tilde{T}^4) = -
        \begin{pmatrix}
            1 & 0\\
            0 & -1\\
        \end{pmatrix},\quad
        \rho_{0,1}(\tilde{U}^4,\tilde{U}) = ie^{3\pi i/4}
        \begin{pmatrix}
            0 & 1\\
            1 & 0\\
        \end{pmatrix}
        ,
    \end{align}
    and
    \begin{align}
        \rho_{0,1}(\tilde{S}^2,\tilde{S}^2) = -1,\quad 
        \rho_{0,1}(\tilde{T},\tilde{T}^4) = -e^{\pi i/4},\quad
        \rho_{0,1}(\tilde{U}^4,\tilde{U}) = ie^{3\pi i/4}.
    \end{align}
    The former is the doublet of $\mathbb{Z}_8\rtimes\mathbb{Z}_2$, while the latter is the $\mathbb{Z}_8$ group.

    For $N=5$, the generators
\begin{align}
    & \rho_{0,1}(\tilde{T},\tilde{T}^5) = -e^{\pi i/4}
    \begin{pmatrix}
        1 & 0 & 0\\
        0 & e^{\pi i/4} & 0\\
        0 & 0 & -1\\
    \end{pmatrix},\quad
    \rho_{0,1}(\tilde{U}^5,\tilde{U}) = \frac{1}{2}
    \begin{pmatrix}
        -1 & \sqrt{2}e^{\pi i/4} & 1\\
        \sqrt{2}e^{\pi i/4} & 0 & \sqrt{2}e^{\pi i/4}\\
        1 & \sqrt{2}e^{\pi i/4} & -1\
    \end{pmatrix},\notag\\
    &\rho_{0,1}(\tilde{U}^2\tilde{S}^3\tilde{U}^3,\tilde{T}^3\tilde{S}^3\tilde{T}^2) = \frac{e^{\pi i/4}}{2}
    \begin{pmatrix}
        i & -\sqrt{2} & i\\
        \sqrt{2}e^{\pi i/4} & 0 & -\sqrt{2}e^{\pi i/4}\\
        -i & -\sqrt{2} & -i\\
    \end{pmatrix},
\end{align}
generate the flavor group $\Delta(48)\rtimes (\mathbb{Z}_2\times \mathbb{Z}_2)$.
The results are summarized in Table \ref{M_1=4,even,M_2=4,odd}.
When $N=3$ and 5, we obtain the same flavor group 
$\Delta(48)\rtimes (\mathbb{Z}_2\times \mathbb{Z}_2)$.
Also, we obtain the same group $\Delta(48)\rtimes (\mathbb{Z}_2\times \mathbb{Z}_2)$ for $N=1$.
When $N=2$ and 4, the flavor groups become smaller, but they include non-Abelian groups.

\begin{table}[H]
    \centering
    \begin{tabular}{|c|c|c|}
    \hline
         $N$ & order& flavor structure  \\
         \hline
         \hline
         $2$&16,8 & doublet of $Q_8$ and singlet of $\mathbb{Z}_8$ \\
         \hline
         $3$& 192& triplet of $\Delta(48)\rtimes (\mathbb{Z}_2\times \mathbb{Z}_2)$\\
         \hline
         $4$&16,8 & doublet of $\mathbb{Z}_8\rtimes\mathbb{Z}_2$ and  singlet of $\mathbb{Z}_8$  \\
         \hline
         $5$& 192& triplet of $\Delta(48)\rtimes (\mathbb{Z}_2\times \mathbb{Z}_2)$\\
         \hline
    \end{tabular}
    \caption{The flavor groups of three zero-modes for $M_1=4$ and $M_2=4$.
    In these models, we take three even modes on $T^2_1/\mathbb{Z}_2$ and the single odd mode on $T^2_2/\mathbb{Z}_2$. 
    For $N=2$ and 4, the three zero-modes are decomposed to a doublet and singlet. The flavor groups of the doublet and singlet are shown.}
    \label{M_1=4,even,M_2=4,odd}
\end{table}

Also, we can study the models with $M_1=M_2=4$ and $\tau_2=N\tau_1$, where 
the single odd mode appears on $T^2_1/\mathbb{Z}_2$ 
and the three even zero-modes appear on $T^2_2/\mathbb{Z}_2$.
These models also lead to three generations, and we obtain the same flavor groups as the ones in Table \ref{M_1=4,even,M_2=4,odd}.

As next examples, we study the models with 
$M_1=4$ and $M_2=2$ and $\tau_2=N\tau_1$.
Here we take the single odd zero-modes on $T^2_1/\mathbb{Z}_2$, and two even zero-modes on $T^2_2/\mathbb{Z}_2$.
Totally, there are two zero-modes. 
The analysis is similar to the above.
Table \ref{M_1=4,odd,M_2=2,even} shows the flavor groups in these models.
When $N=3$ and 5, we obtain the same group 
$T'\rtimes\mathbb{Z}_2$.
We also obtain the same group $T'\rtimes\mathbb{Z}_2$ for $N=1$.
For $N=2$, the flavor group is smaller, but 
non-Abelian.
For $N=4$, the flavor group is much smaller, i.e. Abelian.

\begin{table}[H]
    \centering
    \begin{tabular}{|c|c|c|}
    \hline
    $N$ & order& flavor group  \\
    \hline
    \hline
    $2$& 32& $\Sigma(32)$\\
    \hline
    $3$& 48& $T'\rtimes\mathbb{Z}_2$\\
    \hline
    $4$& 32& $\mathbb{Z}_8\times\mathbb{Z}_4$\\
    \hline
    $5$& 48& $T'\rtimes\mathbb{Z}_2$\\
    \hline
    \end{tabular}
    \caption{The flavor groups of two zero-modes for $M_1=4$ and $M_2=2$.
    In these models, we take the single odd modes on $T^2_1/\mathbb{Z}_2$, and two even modes on $T^2_2/\mathbb{Z}_2$. Totally, there are two zero-modes in these models.}
    \label{M_1=4,odd,M_2=2,even}
\end{table}

As third examples, we examine the models with $M_1=2$ and $M_2=4$ and $\tau_2 = N \tau_1$, 
where the two even modes appear on $T^2_1/\mathbb{Z}_2$, and the single odd mode appears on $T^2_2/\mathbb{Z}_2$.
In these models, we obtain the same flavor groups as the ones in Table \ref{M_1=4,odd,M_2=2,even}.

As further examples, we study the models with $M_1=6$ and 
$M_2=2$ and $\tau_2=N\tau_1$.
Here, we take the odd zero-modes on $T^2_1/\mathbb{Z}_2$, 
and their number is two.
On $T^2_2/\mathbb{Z}_2$, there are two even zero-modes.
Totally, there are four zero-modes. 
The analysis is similar to the above.
Table \ref{M_1=6,odd,M_2=2,even} shows the flavor groups of four zero-modes in these models.
When $N=3$ and 5, four zero-modes are decomposed into a triplet and a singlet.
The flavor group of the triplet is 
$S_4\times (\mathbb{Z}_3\times \mathbb{Z}_2)$, while the flavor group of the singlet is $ \mathbb{Z}_3\times \mathbb{Z}_2$.
These groups are the same for $N=3$ and 5.
The flavor group $S_4\times (\mathbb{Z}_3\times \mathbb{Z}_2)$ is larger than the flavor group for $N=1$, while $N=1$ leads to the flavor group 
$A_4\times (\mathbb{Z}_3\times \mathbb{Z}_2)$.
When $N=2$ and $N=4$, the flavor groups become smaller.
In particular, the flavor group for $N=4$ is Abelian.

\begin{table}[H]
    \centering
    \begin{tabular}{|c|c|c|}
    \hline
         $N$ & order& flavor structure  \\
         \hline
         \hline
         $2$& 96& quartet of $(\mathbb{Z}_4\times\mathbb{Z}_3\times\mathbb{Z}_2)\rtimes\mathbb{Z}_4$\\
         \hline
         $3$&144, 6 & triplet of $S_4\times\mathbb{Z}_3\times\mathbb{Z}_2$ and and singlet of $\mathbb{Z}_3\times\mathbb{Z}_2$ \\
         \hline
         $4$& 48&  four singlets of $\mathbb{Z}_4\times\mathbb{Z}_4\times\mathbb{Z}_3$\\
         \hline
         $5$&144, 6 & triplet of $S_4\times\mathbb{Z}_3\times\mathbb{Z}_2$ and singlet of $\mathbb{Z}_3\times\mathbb{Z}_2$ \\
         \hline
    \end{tabular}
    \caption{The flavor groups of three zero-modes for $M_1=6$ and $M_2=2$.
    In these models, we take odd modes on $T^2_1/\mathbb{Z}_2$, whose number is two, and even modes on $T^2_2/\mathbb{Z}_2$, whose number is two. Totally, there are four zero-modes in these models. For $N=3$ and 5, the four zero-modes are decomposed into a triplet and a singlet. Their flavor groups are shown.}
    \label{M_1=6,odd,M_2=2,even}
\end{table}

Also, we can study the models with $M_1=2$ $M_2=6$ and $\tau_2=N\tau_1$, where 
the two even mode appears on $T^2_1/\mathbb{Z}_2$ 
and the two odd zero-modes appear on $T^2_2/\mathbb{Z}_2$.
These models also lead to four zero-modes.
In these models, we obtain the same flavor groups as the ones in Table \ref{M_1=6,odd,M_2=2,even}.

Similarly, we can discuss models with other values of $M_1$ and $M_2$ and large values of $N$.
We can realize various flavor groups.
We may have constraints on larger values of $N$ as well as $M_1$ and $M_2$.
For example, values of $N$ are determined by 
three-form fluxes, 
and larger values are constrained by 
the Ramond-Ramond (RR) tadpole cancellation condition \cite{Blumenhagen:2006ci,Ibanez:2012zz}.
Note that the three-form fluxes and magnetic fluxes, $M_{1,2}$ contribute to the RR-charges.
To examine explicit constraints, 
we have to study concrete models.
That is important but beyond our scope.
We would study it elsewhere.


\section{Conclusion}
\label{sec:conclusion}

We have studied the flavor symmetries of zero-modes, which are originated 
from the modular symmetry on $T^2_1\times T^2_2$ and its orbifold with magnetic fluxes.
Our models may correspond to the effective field theory of magnetized $D7$-brane models.
We have introduced the constraint on the moduli parameters by $\tau_2=N\tau_1$.
Such a constraint can be realized by the moduli stabilization due to three-form flux backgrounds.
This constraint breaks the modular symmetry $SL(2,\mathbb{Z})_{\tau_1}\times SL(2,\mathbb{Z})_{\tau_2}$ of two moduli, $\tau_1$ and $\tau_2$ to 
$\Gamma_0(N) \times \Gamma^0(N)$.
We can derive various flavor symmetries of zero-modes.
Some of them are the same as the group for $N=1$, but the cases with $N=2$ and 4 
lead to different groups, which are non-Abelian in many cases and interesting.
Abelian flavor groups are also interesting.
(See e.g. Ref.~\cite{Kikuchi:2023fpl}.)
It would be interesting to build flavor models using these groups.
We leave it as a future work.

We have started with $SL(2,\mathbb{Z})_{\tau_1}\times SL(2,\mathbb{Z})_{\tau_2}$.
Then, we have introduced the constraint $\tau_2=N\tau_1$ by physical reason, which is coming from the moduli stabilization. 
As a result, we have obtained various interesting finite flavor groups. 
It is interesting to extend our analysis to the cases including various constraints such as 
$N_2 \tau_2=N_1\tau_1$ for co-prime integers, 
$N_1$ and $N_2$.
It is also important to study 
more than two moduli parameters.
We would study them elsewhere.


\vspace{1.5 cm}
\noindent
{\large\bf Acknowledgement}\\

This work was supported in part JSPS KAKENHI Grant Numbers JP23K03375 (T.K.), JP24KJ0249 (K.N.), and JP23H04512 (H.O), and JST SPRING Grant Number JPMJSP2119(S.T.).

\appendix
\section{Computation of $\rho(\tilde{U}^N)$}
\label{app}


Here, we show that 

\begin{align}
        \rho(\tilde{U}^N)_{jk} = \frac{(-1)^{N-1}i}{\sqrt{NM}}e^{\pi i/4}e^{\frac{\pi i}{NM}(j-k)^2}\sum^{N-1}_{l = 0}e^{\frac{\pi i M}{N}l^2}e^{-\frac{2\pi i}{N}(j-k)l}.
    \end{align}
    \begin{proof}
  Let us show the above relation by mathematical induction.\\
When $N=1$, it is straightforward to derive the above form by calculating directly from $\rho(\tilde{S}),\rho(\tilde{T})$.\\
Next, we assume that 
\begin{align}
  \rho(\tilde{U}^N)_{jk} = \frac{(-1)^{N-1}i}{\sqrt{NM}}e^{\pi i/4}e^{\frac{\pi i}{NM}(j-k)^2}\sum^{N-1}_{l = 0}e^{\frac{\pi i M}{N}l^2}e^{-\frac{2\pi i}{N}(j-k)l},
\end{align}
for any $N$.
Then, we compute $\rho(\tilde{U}^{N+1})$ as 
\begin{align}
  \rho(\tilde{U}^{N+1})_{jk} &= \sum^{M-1}_{l=0}\rho(\tilde{U})_{jl}\rho(\tilde{U}^N)_{lk} \notag \\
  &=\frac{(-1)^Ni}{\sqrt{N}M}\sum^{M-1}_{l=0}e^{\frac{\pi i}{M}(j-l)^2}e^{\frac{\pi i}{NM}(l-k)^2}\sum^{N-1}_{m=0}e^{\frac{\pi iM}{N}m^2}e^{\frac{2\pi i}{N}(-l+k)m} \notag \\
  &= \frac{(-1)^Ni}{\sqrt{N}M}e^{\frac{\pi i}{NM}(Nj^2+k^2)}\sum^{M-1}_{l=0}e^{\frac{\pi i}{NM}\{ (N+1)l^2-2(Nj+k)l \}}\sum^{N-1}_{m=0}e^{\frac{\pi iM}{N}m^2}e^{\frac{2\pi i}{N}(-l+k)m}.
\end{align}
The summations in the right hand side is slightly complicated.
We define $A$ as 
\begin{align}
  A = \sum^{M-1}_{l=0}e^{\frac{\pi i}{NM}\{ (N+1)l^2-2(Nj+k)l \}}.
\end{align}
Then, $A$ can be decomposed as 
\begin{align}
  A &= \sum^{NM-1}_{l=0}e^{\frac{\pi i}{NM}\{ (N+1)l^2-2(Nj+k)l \}}-\sum^{N-1}_{s=1}\sum^{(s+1)M-1}_{l=sM}e^{\frac{\pi i}{NM}\{ (N+1)l^2-2(Nj+k)l \}} \notag \\
  &= \sum^{NM-1}_{l=0}e^{\frac{\pi i}{NM}\{ (N+1)l^2-2(Nj+k)l \}}-\sum^{N-1}_{s=1}\sum^{M-1}_{l=0}e^{\frac{\pi i}{NM}\{ (N+1)l^2-2(Nj+k)l \}}e^{\frac{\pi iM}{N}s^2}e^{\frac{2\pi i}{N}(-l+k)s}.
\end{align}
That results in 
\begin{align}
   \sum^{NM-1}_{l=0}e^{\frac{\pi i}{NM}\{ (N+1)l^2-2(Nj+k)l \}}&=
   A+\sum^{N-1}_{s=1}\sum^{M-1}_{l=0}e^{\frac{\pi i}{NM}\{ (N+1)l^2-2(Nj+k)l \}}e^{\frac{\pi iM}{N}s^2}e^{\frac{2\pi i}{N}(-l+k)s} \notag \\
   &=\sum^{N-1}_{s=0}\sum^{M-1}_{l=0}e^{\frac{\pi i}{NM}\{ (N+1)l^2-2(Nj+k)l \}}e^{\frac{\pi iM}{N}s^2}e^{\frac{2\pi i}{N}(-l+k)s}.
\end{align}
By use of this, we can write 
\begin{align}
  \rho(\tilde{U}^{N+1})_{jk} = \frac{(-1)^Ni}{\sqrt{N}M}e^{\frac{\pi i}{NM}(Nj^2+k^2)}\sum^{NM-1}_{l=0}e^{\frac{\pi i}{NM}\{ (N+1)l^2-2(Nj+k)l \}}.
\end{align}
We can calculate this summation by using the following relation called Landsberg–Schaar relation:
\begin{align}
    L(a,b,c) = \sqrt{\Big|\frac{c}{a}\Big|}\exp\Big\{\frac{\pi i}{4}\Big({\rm sgn}(ac)-\frac{b^2}{ac}\Big)\Big\}L(-c,-b,a),
  \end{align}
with
  \begin{align}
    L(a,b,c) = \sum^{|c|-1}_{n=0}\exp{\frac{\pi i(an^2+bn)}{c}},
  \end{align}
  where $a, b, c \in \mathbb{Z}$, $a, c \neq 0$, $ac+b \equiv 0\pmod 2$, and
  \begin{align}
      {\rm sgn}(ac) = 
      \left\{ \,
      \begin{aligned}
          &+1\quad (ac > 0)\\
          &-1\quad (ac < 0)\\
      \end{aligned}
      \right.
      .
  \end{align}
Then $\rho(\tilde{U}^{N+1})_{jk}$ becomes
\begin{align}
  \rho(\tilde{U}^{N+1})_{jk} &= \frac{(-1)^Ni}{\sqrt{N}M}e^{\frac{\pi i}{NM}(Nj^2+k^2)}\sqrt{\frac{NM}{N+1}}e^{\pi i/4}e^{-\frac{(Nj+k)^2}{N(N+1)M}\pi i}\sum^{(N+1)-1}_{l=0}e^{\frac{\pi iM}{N+1}l^2}e^{-\frac{2\pi i}{N+1}(j-k)l}\nonumber\\
  &= \frac{(-1)^{(N+1)-1}i}{\sqrt{(N+1)M}}e^{\pi i/4}e^{\frac{\pi i}{(N+1)M}(j-k)^2}\sum^{(N+1)-1}_{l=0}e^{\frac{\pi iM}{N+1}l^2}e^{-\frac{2\pi i}{N+1}(j-k)l}.
\end{align}
\end{proof}

\clearpage

\end{document}